\documentclass[10pt,twocolumn]{article}

\usepackage{amsmath}
\usepackage{graphicx}
\usepackage{color}
\usepackage{rotating}

\usepackage[round,numbers,sort&compress]{natbib} 
\DeclareTextSymbol{\degre}{OT1}{23} 
\newcommand{\latin}[1]{#1} 

\newcommand{\rumycommentold}[1]{} 

\newcommand{\corr}[1]{#1} 
\newcommand{\corrbis}[1]{{#1}} 

\title{\LARGE \textbf{A New Method for Measuring Edge Tensions \corr{and Stability of Lipid Bilayers}:\\ \LARGE Effect of Membrane Composition}}

\author{Thomas~Portet \\
    \small Institut de Pharmacologie et de Biologie Structurale, CNRS UMR 
5089;\\ \small Laboratoire de Physique Th\'eorique, CNRS UMR 5152;\\ 
\small Universit\'e Paul Sabatier, Toulouse, France.
    \and Rumiana~Dimova \thanks{
           Corresponding author.  Address: 
           Max Planck Institute of Colloids and Interfaces,
       Science Park Golm,
       14424 Potsdam, Germany.
       Tel.:~+49 331 567 9615, Fax:~+49 331 567 9612, 
E-mail:~Rumiana.Dimova@mpikg.mpg.de}\\
    \small Max Planck Institute of Colloids and Interfaces, \\
    \small Potsdam, Germany.}	

\date{}

\begin{document}

\maketitle


\abstract{We report a new and facile method for measuring edge tensions of lipid membranes. The approach is based on electroporation of giant unilamellar vesicles and analysis of the pore closure dynamics. We applied this method to evaluate the edge tension in membranes with four different compositions: egg phosphatidylcholine (EggPC), dioleoylphosphatidylcholine (DOPC), and mixtures of the latter with cholesterol and dioleoylphosphati\-dylethanolamine (DOPE). Our data confirm previous results for EggPC and DOPC. The addition of 17 mol \% cholesterol to the DOPC membrane causes an increase in the membrane edge tension. On the contrary, when the same fraction of DOPE is added to the membrane, a decrease in the edge tension is observed, which is an unexpected result considering the inverted-cone shape geometry of the molecule. \corr{Presumably, interlipid hydrogen bonding lies in the origin of this behavior.} Furthermore, cholesterol was found to lower the lysis tension of DOPC bilayers. This behavior differs from that observed on bilayers made of stearoyloleoylphosphatidylcholine, suggesting that cholesterol influences the membrane mechanical stability in a lipid-specific manner. 
}


\section*{Introduction}
In order to fulfill the role of a barrier, spontaneous pore formation is suppressed in biological membranes. When a membrane is porated due to some external perturbation, lipid molecules reorient so that their polar heads can line the pore walls and form a hydrophilic pore \citep{Lit75}. This rearrangement is energetically favorable because it shields the hydrophobic tails from exposure to water. However, there is still a price to pay for reorganizing the lipids at the pore edge. The energy penalty per unit length of pore circumference is called edge tension and is on the order of several picoNewtons \cite{Boa02}. The edge tension emerges from the physicochemical properties and the amphiphilic nature of lipid molecules, and gives rise to a force driving the closure of transient pores \citep{Bro00}.

In some reports, the edge tension is also referred to as line tension. The latter term is presumably more appropriate for describing, for example, the tension along the borders of domains in multicomponent membranes. In porated bilayers, we will refer to the tension along the pore edge as edge tension.

The edge tension is of significant importance in biology. It contributes to the auto-assembly and auto-healing of lipid bilayer structures, which enable compartmentalization in cells needed for life to develop. Edge tension has been found to govern the stability of detergent-stabilized bicelles and to regulate the disc-to-vesicle transitions in such systems \citep{Fro86}. It also plays a crucial role in membrane resealing mechanisms taking place after physical protocols for drug delivery, such as  sonoporation \citep{New07} or electroporation \citep{Esc09}. Being able to experimentally measure this quantity is thus of significant interest for understanding various biological events and physicochemical processes in membranes. Studies based on molecular dynamics simulation have also been able to deduce membrane edge tensions \citep{Leo04,Jia04,Tol04}.

Only a few experimental methods have been developed to directly assess the membrane edge tension. In the following, we give a short overview of the available approaches and their advantages, and briefly outline the disadvantages. One of the main hurdles in measuring the edge tension lies in visualizing the pores in the membrane. Indirect methods based on measuring properties statistically averaged over populations of small vesicles provide a possible solution, as reported, for example, in studies on osmotically-induced leakage of vesicles \cite{Tau75}. However, in such measurements the use of highly concentrated salt solutions influences significantly the membrane material properties \cite{Pabst07}. \corrbis{Another} approach was based on rapid freezing of cells with a controlled time delay after electroporation, and examining the pores with electron microscopy \cite{Chang90}. This method, however, provides a static picture of the porated membrane and a danger of ice crystals damage exists. Another way to probe the edge energy, this time on supported lipid bilayers, is provided by a ``punch-through'' approach with an atomic force microscope (AFM) \citep{Loi02}, but here, the support may influence the membrane behavior. Measurements of the voltage dependence of the average lifetime of pores in electroporated black lipid membranes also can be used to roughly estimate the edge tension \citep{Mel90,Gen93}. However, these membranes are at high tension and organic solvents may still be present in the bilayer.  

The above listed approaches and systems do not allow access to the pore dynamics and the pores cannot be directly observed. A more convenient system, which provides solution to this problem is giant unilamellar vesicles (GUVs) \cite{Luisi00}. Having the dimensions of eukaryotic cells, GUVs can be visualized under an optical microscope and present a convenient tool for studying membrane properties \citep{Dim06}.

Only a few previous studies have employed GUVs for estimating the edge tension. Observations on open cylindrical giant vesicles exposed to AC fields \citep{Har79} provided an estimate for the edge tension, but this technique could be applied to only few (three) vesicles. In  another work, the vesicles were porated with an electric pulse, and the pores were kept open by externally adjusting the membrane tension with a micropipette \citep{Zhe93}. This approach \corrbis{nevertheless} requires the use of a sophisticated equipment like a setup for vesicle micropipette aspiration. Giant liposomes were also used in \citep{San99,Kar03}, where the pore closure dynamics was analyzed in light of a theory developed earlier \citep{Bro00}. However, for the direct visualization of the pore closure, the use of viscous media (glycerol solutions) and fluorescent dyes in the membrane were required, both of which potentially influence the edge tension.  
\corrbis{Indeed, glycerol is known to interact with phosphatidylcholine membranes \latin{via} preferential exclusion from the hydration layer and membrane partitioning \citep{Wes03}. It is thus conceivable that glycerol can influence the edge tension, because the latter depends on the repulsion between the lipid headgroups and the conformation of the hydrocarbon chains as shown by simulation studies \citep{May00}. On the other hand, the use of fluorescence dyes as in \citep{Kar03} to visualize the vesicles can also influence the values of the measured edge tension because of preferential partitioning of the dye between the pore edges and the rest of the bilayer.}

Similarly to \citep{San99,Kar03}, we also use the theoretical framework developed by Brochard-Wyart et al. \citep{Bro00} describing pore dynamics. The principle of our method is simple. One has to create a pore and measure its closure rate, which can be related to the edge tension following \citep{Bro00}. Porating a lipid bilayer can be carried out by applying tension to the membrane. Once the membrane tension exceeds a critical value called the lysis tension \citep{Boa02}, the vesicle ruptures and a transient pore can appear. Karatekin~\latin{et~al.} used visible light illumination to stretch the vesicles \citep{Kar03}. However, when working with fluorescently labeled membranes as in \citep{Kar03}, the danger exists that intense illumination may trigger oxidation processes in the bilayer \citep{Auyuan06,Zhao07}. Sandre et al. employed adhesion of the GUVs to a glass substrate \citep{San99}, but this approach does not allow control over the \corrbis{location and initiation} of the poration process.

Here, as in \citep{Zhe93}, we use an electric pulse to generate micron-sized pores in the membrane. 
We observe the pore closure under phase contrast microscopy using fast digital imaging. Thus, the need of using viscous solutions to slow down the system dynamics was avoided and no fluorescent dyes to visualize the vesicles were employed. The pores close within a few tens or hundreds of milliseconds, giving edge tensions of the order of several picoNewtons, in agreement with values reported in the literature. We study the influence of the membrane composition and investigate effects resulting from the inclusion of cholesterol or another lipid type. In addition, we were able to evaluate the threshold values of the transmembrane potential needed to porate bilayers with various compositions and summarize these results in terms of lysis tensions at the end. 


\section*{Method description}
\subsection*{Theoretical considerations of pore dynamics}

According to \citep{Bro00}, the life of a pore \corrbis{in a spherical vesicle of radius $R$} is composed of four consecutive stages. Initially, the pore grows and then stabilizes to its maximal radius. After that, the pore radius decreases slowly in a quasi-static leakout regime, followed by the last stage of fast closure. The pore spends the majority of its lifetime in the third regime of slow closure, which we use to determine the edge tension. \corrbis{For a circular pore of radius $r$}, it can be shown, see Supporting Material, that 

\begin{equation}
R^2 \ln (r)=-\frac{2\gamma}{3\pi\eta}t + \mbox{$C$} \ ,
\label{eqn:R2lnr}
\end{equation}
where $\gamma$ denotes the edge tension, $\eta$ is the viscosity of the aqueous medium, $t$ is time, and $C$ \corrbis{is a time-independent constant taking one value for each experiment.} \corrbis{This equation is obtained from combining three coupled differential equations for the three unknowns $R$, $r$ and the vesicle surface tension $\sigma$, using the approximation that $r$ changes slowly (i.e. $r\sigma \approx\gamma$, see Supporting Material). This final equation is solved assuming constant $R$.} 

The principle of the edge tension measurement is very simple. One only has to consider the linear part of $R^2 \ln (r)$ as a function of time corresponding to the slow closure stage. Linear fit of this part gives a slope $a$ and the edge tension $\gamma$ is estimated from the relation $\gamma=-(3/2)\pi\eta a$.

\subsection*{Electromediated pore formation}

We trigger the process of electroporation by  applying an electric pulse of 5 ms duration and field strength in the range 20 - 80 kV/m. The pulse leads to a voltage drop across the membrane of the order of 1~V. Such pulses can create pores several microns in diameter, also referred to as macropores. We monitor the pore size evolution with fast phase contrast imaging.
\corrbis{In this way we avoid the use of a fluorescence dye or glycerol as in \citep{Kar03} and expect to obtain more accurate values for the edge tension.}. 

Using phase contrast microscopy, it is possible to accurately measure the size of the pore if located in the in-focus area of the vesicle as has been demonstrated in earlier electroporation and fusion studies \citep{Riske05,Hal06}. Here, the pores were always located in this area, and more precisely at the vesicle poles facing the electrodes, because of the angular dependence of the transmembrane potential created during the pulse and the chamber geometry; see e.g. \citep{Dimova07} and Supporting Material. We also employed a conventional trick to improve the contrast in the vesicle images: we used different sugar solutions inside and outside the liposomes. The vesicles were prepared in a sucrose solution and subsequently diluted in a glucose solution. Due to the difference in the refractive indices of these solutions, the vesicles appear as dark objects surrounded by a bright halo on a light-gray background; see the phase contrast image in Fig.~\ref{fig:photo}A. The vesicle membrane is located where the image intensity gradient is maximal. When a pore opens, the two sugar solutions mix and the gradient vanishes in the pore area. Thus, interrupting the membrane continuity in the focal plane, i.e. inducing pores in this area, causes an interruption in the halo, which we used to determine the pore radius. We developed a program to automatically measure the pore sizes in order to avoid any bias introduced by manual processing; see Image analysis subsection and Supporting Material for details.

Let us note that many small suboptical pores located very close to each other could lead to images similar to those obtained for a single big pore. Furthermore, in some cases several macropores could be created within the same vesicle, occasionally on both sides facing the electrodes. In this case, Eq.~\ref{eqn:R2lnr} is still valid; see Supporting Material for justifications.

The overall vesicle response and behavior of the pores over time as theoretically described in \citep{Bro00} is different from the one observed here in the stage of pore formation. The applied electric field leads to a gradual and nonuniform increase of the membrane tension along the vesicle surface. Thus, the theoretical approach in \citep{Bro00}, which assumes constant and uniform tension distribution all over the vesicle, cannot be applied to the period during the pulse. \corr{However, after the end of the pulse, the membrane tension is no longer inhomogeneous and the pore closure analysis are applicable (as detailed in the Supporting Material).}

During the pulse when the tension exceeds some critical value corresponding to the lysis tension, the membrane ruptures, i.e., a pore is formed. Membrane electroporation is associated with building of some critical transmembrane potential across the bilayer, \corr{which we use to evaluate the lysis tension of membranes of various composition}. 


\section*{Materials and experimental procedures}
\subsection*{Lipids and solutions}
Egg yolk L-$\alpha$-phosphatidylcholine (EggPC), dioleoylphosphatidylcholine\newline (DOPC), dioleoylphosphatidylethanolamine (DOPE) and L-$\alpha$-phosphatidyl-ethanolamine-N-(lissamine rhodamine B sulfonyl) (Rhodamine PE) were purchased from Avanti Polar Lipids (Alabaster, AL), and cholesterol from Sigma (Steinheim, Germany). Four different membrane compositions were studied: DOPC, DOPC:cholesterol at a molar ratio of 5:1, DOPC:DOPE, also at a molar ratio of 5:1, and EggPC fluorescently labeled with Rhodamine PE (1 mol \%). Lipids were diluted in chloroform at a mass concentration of 0.5 mg/mL, and stored at -20\degre C.
The vesicles were prepared in an aqueous solution of 240 mM sucrose (internal solution) and subsequently diluted in an aqueous solution of 260 mM glucose and 1 mM sodium chloride (external solution). The pH of the solutions was adjusted to 7 using either 1 mM phosphate buffer (KH$_2$PO$_4$/K$_2$HPO$_4$, Merck, Darmstadt, Germany) or 1 mM Hepes buffer from Sigma. No difference in the behavior of the vesicles was observed when using one buffering agent or the other. The conductivities of the internal and external solutions were measured with conductivitymeter SevenEasy (Mettler Toledo, Greifensee, Switzerland), and had values of approximately  20 $\mu$S/cm and 150 $\mu$S/cm, respectively. 
The osmolarities, measured with osmometer Osmomat 030 (Gonotec, Berlin, Germany), were approximately 260 mOsm/kg and 275 mOsm/kg, respectively. 

\subsection*{Giant vesicles preparation}
The GUVs were prepared using the electroformation protocol \citep{Ang86} at room temperature, at which all lipids are in the fluid phase. A small volume (15 $\mu$L) of the lipid solution in chloroform was deposited on the conducting sides of glass slides coated with indium tin oxide. The glasses were then kept for two hours under vacuum at 63\degre C in an oven to remove all traces of the organic solvent. Afterwards, the plates, spaced by a 1 mm thick silicon frame (Electron Microscopy Sciences, Hatfield, PA) were assembled into a chamber. The chamber was filled with the sucrose solution (internal medium). The slides were connected to AC field function generator Agilent 33220A (Agilent Technology Deutschland GmbH, B\"oblingen, Germany) and sinusoidal voltage of 25 mV peak to peak at 10 Hz was applied. The voltage was increased by 100 mV steps every 5 minutes, up to a value of 1225 mV and maintained under these conditions overnight. \corr{Shorter times ($\approx2$ h) were also sufficient without altering the results.} Finally, square-wave AC field of the same amplitude was applied at 5 Hz for one hour in order to detach the GUVs from the slides.

\subsection*{Application of electric pulses}
For pulse application, we used a home-made chamber constructed from a glass slide and a coverslip similar to the one described in \citep{Por09}. Two parallel copper strips (3M, Cergy-Pontoise, France) were stuck on the slide, 0.5 cm apart. The coverslip was then glued onto the glass slide with heated parafilm. The cavity between the slide and the coverslip was filled with 30 $\mu$L of the buffered glucose solution and 2 $\mu$L of the GUV solution. Electric pulses of 5 ms duration and amplitudes ranging from 20 to 80 kV/m were applied directly under the microscope with $\beta$tech pulse generator GHT\_Bi500 ($\beta$tech, l'Union, France). The pulse generator was synchronized with the digital camera, so that the image acquisition was triggered by the onset of the pulse. 

\subsection*{Microscopy}
The GUVs were observed under inverted microscope Axiovert 135 (Zeiss, G\"ottingen, Germany) equipped with a 20$\times$ Ph2 objective. Phase contrast images were collected at acquisition speeds of 5000, 10000 or 20000 frames per seconds (fps) with fast digital camera HG-100K (Redlake, San Diego, CA). Illumination during routine observations was performed with a halogen lamp. A mercury lamp HBO W/2 was used for image acquisition for a total time of a few seconds. During this time no detectable heating of the sample occurred. 

\subsection*{Image analysis}
To avoid bias in the pore size determination, we developed image processing procedures to automatically evaluate the pore radii. The procedure consists of two stages: (\textit{i}) detection of the vesicle contour, and (\textit{ii}) pore size measurement. We first transformed the raw vesicle image into a binary one, the membrane being represented by the non zero pixel value. Given such an image in which the decision of where the membrane was located had already been made, it was then easy to obtain the pore size $r$ \latin{via} custom algorithms, see Supporting Material.

The vesicle radii $R$ were measured manually using Image J (National Institutes of Health, Bethesda, MD). As described in the previous section, the linear part of the curve $R^2 \ln (r)$ as a function of time $t$ was fitted with an expression of the form $y=at+b$, and the edge tension $\gamma$ was obtained from the slope $a$ through $\gamma=-(3/2)\pi\eta a$. For the viscosity of the 260 mM glucose solution, we used $\eta = 1.133\ 10^{-3}$ Pa.s \citep{Weast88}. To avoid effects due to change in the vesicle radius, predominantly occurring before the stage of linear dependence of the data, see Supporting Material, $R$ was measured at the end of this stage.


\section*{Results}
\subsection*{Edge tension of EggPC membranes}
\begin{figure*}
   \begin{center}
      \includegraphics*[width=12.5cm]{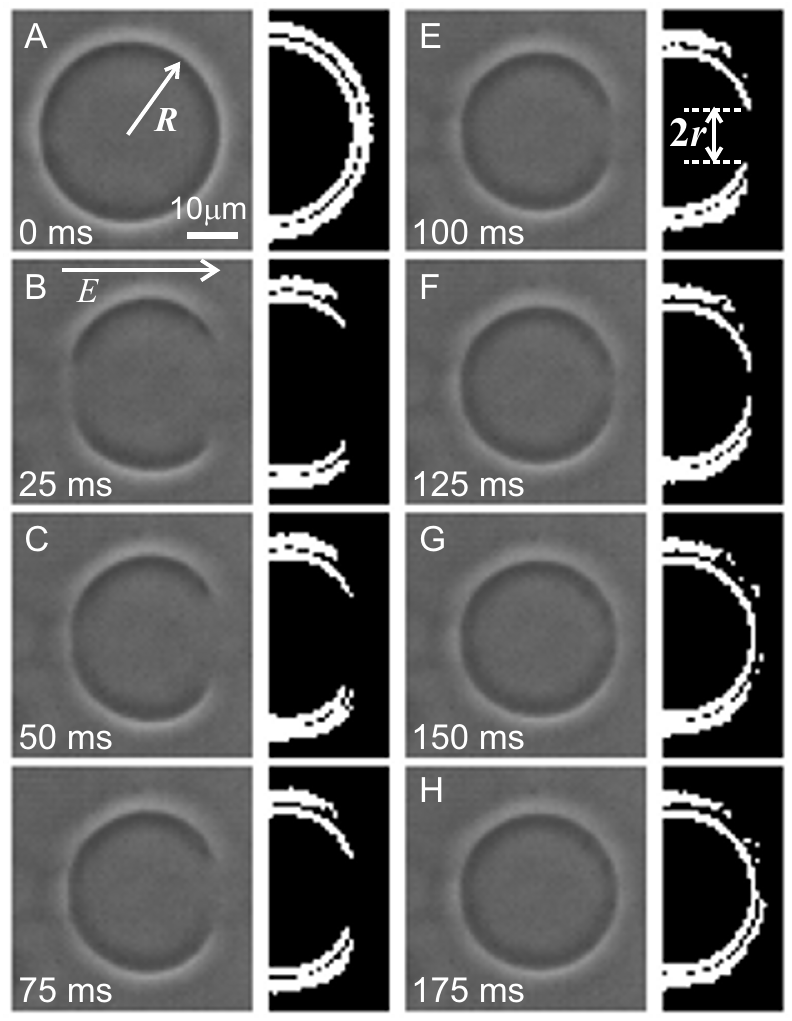}
      \caption{Time sequence of raw images (first and third columns) and processed images (second and fourth columns) of a vesicle with a radius of 17 $\mu$m exposed to an electric pulse of 5 ms duration and 50 kV/m amplitude. Time $t=0$ s (A) corresponds to the beginning of the pulse. The pore radius $r$, schematically indicated in (E), decreases (B-F) and after around 150 ms it has resealed (G, H). The field direction is indicated in (B). A pore formed in the left hemisphere of the vesicle is visible in (B).}
      \label{fig:photo}
   \end{center}
\end{figure*}

Because the majority of literature data on edge tensions has been collected on EggPC membranes, we first tested our method on GUVs made of this lipid. With fluorescence microscopy of EggPC vesicles labeled with Rhodamine PE (data not shown), we confirmed that single macropores formed on the cathode-facing pole of the vesicles. This supports previous observations on DOPC \citep{Tek01,Por09} and indicates that such behavior is not tail-specific but is presumably related to the nature of the lipid headgroup. Only in a limited number of experiments, macropores could be visualized also on the anode-facing side of the vesicles. Thus, the pore sizes were measured only at the cathode-facing hemispheres. For the vesicle radii explored here (between 10 and 40 $\mu$m), the applied pulses induce transmembrane potentials of the order of 0.7--1.25 V. The evolution of the closing pore after the end of the pulse could be clearly followed with phase contrast microscopy as shown with the example in Fig.~\ref{fig:photo}. 
The very first image shows the unperturbed vesicle.


During the pulse, i.e., in the first 5 ms (data not shown), the vesicles attained spherocylindrical shapes with symmetry axis along the field direction (see e.g. Fig.~\ref{fig:burst}B), as previously reported for GUVs in salt solutions \citep{Ris06}. The deformation here seemed less pronounced, presumably because of the lower field amplitude used, but with a longer lifetime, on the order of a few ms as compared to hundred $\mu$s in \citep{Ris06}, because of the longer pulses applied in this work.

The pores could only be detected after about 5 ms, almost at the end of the pulse. The temporal and optical resolution of our setup did not allow us to detect the pore formation and growth precisely. It appeared that the pore was created already with a rather large radius of a few micrometers, slightly grew in the following few hundreds of microseconds, stabilized in the next couple of milliseconds and then entered the stage of slow closure. This sequence was exhibited by almost every vesicle we studied, for all compositions. It remains unclear whether the poration process was initiated with a single pore created during the pulse, or that first many smaller pores were formed and then coalesced. If the first scenario is valid, the rate of opening of the macropore should be on the order of several mm/s, i.e., beyond our temporal and spatial resolution. We favor the second hypothesis with smaller pores of suboptical sizes since the critical transmembrane potential for poration is reached already in the beginning of the pulse; see Supporting Material.

We now focus on the stage of pore closure used to determine the edge tension. The images in Fig.~\ref{fig:photo}B-F belong to this stage. The pore has completely resealed after 150 ms; see Fig.~\ref{fig:photo}G, H. 
The image at 25 ms, Fig.~\ref{fig:photo}B, shows that the GUV membrane is altered at the anode-facing side as well. This hemisphere has been reported to be the location of smaller pores nucleation \citep{Tek01} and generation of membrane tubes \citep{Por09}, which apart from the fact that poration was not always observed, was an argument against analyzing the pore closure at this hemisphere. Here the resealing occurred much faster: this side of the liposome is intact on the subsequent pictures. 
To illustrate the image analysis procedure leading to the quantitative measurement of the pore radius, see also Supporting Material, in the second and fourth columns of Fig.~\ref{fig:photo} we present the processed images corresponding to the cathode-facing half of the GUV shown in the adjacent raw images. The outer white circle in the processed image corresponds to the boundary between the bright halo around the vesicle and the gray background, while the inner one indicates the membrane position. It is this inner circle, which is used for pore size measurements. The pore diameter is schematically indicated in Fig.~\ref{fig:photo}E.

The electroporated GUVs generally decreased in size upon the application of the pulse (as exemplified by the vesicle in Fig.~\ref{fig:photo}). This behavior is consistent with previous reports \citep{Por09}, but questions the assumption for constant $R$, which was used to derive Eq.~\ref{eqn:R2lnr}. However, the change in the vesicle radius, probably linked to the expulsion of small vesicles or the pulling of membrane tubes \cite{Por09}, occurs predominantly in the first hundreds of microseconds after the pulse, followed by a negligible decrease of approximately 1.5 \% during the stage of pore closure; see Supporting Material. The latter change is on the order of the error in measuring the pore radius. Thus, our assumption for constant vesicle size in the linear stage of slow pore closure is justified. For the value of $R$ in the fits according to Eq.~\ref{eqn:R2lnr} we used the liposome radius after the pore has closed.

\begin{figure*}
   \begin{center}
      \includegraphics*[width=16cm]{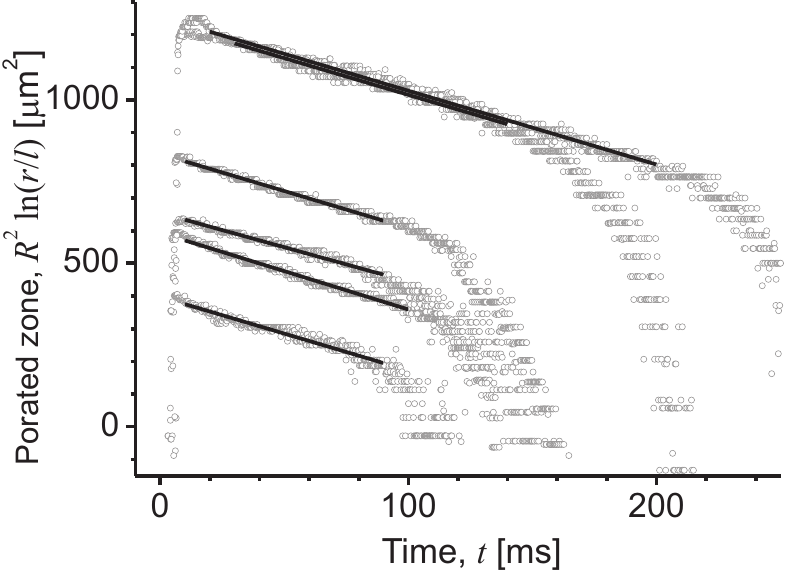}
      \caption{Evolution of the porated region as characterized by $R^2 \ln (r/l)$ as a function of time $t$ for 6 different EggPC vesicles. \corr{To avoid plotting a dimensional value in the logarithmic term, we have introduced $l=1\ \mu$m (note that this operation does not influence the slope of the curves)}. The gray open circles are experimental data and the solid lines are linear fits, whose slopes yield the edge tension $\gamma$. \corrbis{The vesicle radii $R$ are, from bottom curve to top curve: 12.7~$\mu$m,   16.0~$\mu$m,   17.6~$\mu$m,   21.1~$\mu$m,   20.7~$\mu$m and 21.7~$\mu$m.}}
      \label{fig:figure2}
   \end{center}
\end{figure*}

\begin{figure*}
   \begin{center}
      \includegraphics*[width=16cm]{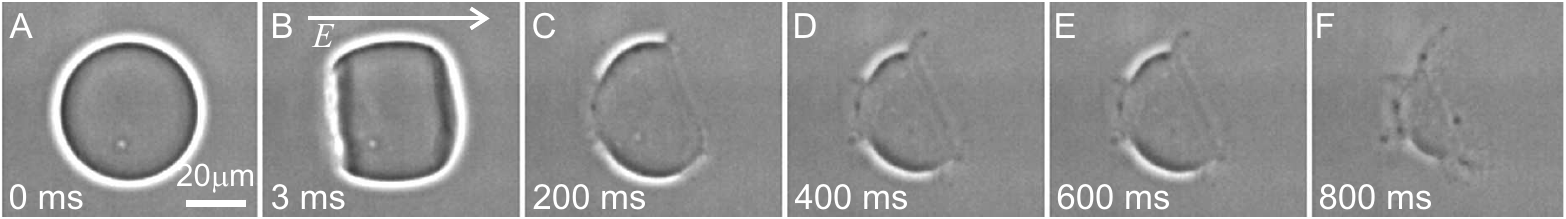}
      \caption{Image sequence of a collapsing vesicle made of DOPC:cholesterol 5:1 exposed to a pulse with a field strength of 50 kV/m and a duration of 5 ms. The field direction is indicated in (B) where the vesicle attains a spherocylindrical shape.}
      \label{fig:burst}
   \end{center}
\end{figure*}

From the images, we extract the time dependence of the pore size. We plot the size of the porated regions characterized by $R^2 \ln (r)$ as a function of time. In Fig.~\ref{fig:figure2} we show 6 typical datasets from the 41 measurements performed on EggPC GUVs. \corrbis{The linearity of the time dependence of the quantity $R^2\ln (r)$ in the quasi-static leakout regime and the similar slopes in the different datasets emphasize the validity of the theoretical model.} The straight solid lines are fits according to Eq.~\ref{eqn:R2lnr}. The edge tension $\gamma$ is deduced from the slopes of these lines. The slopes are similar, because the explored vesicles have identical compositions, i.e., identical edge tension. For EggPC membranes, we found that the mean value of the edge tension was 14.2 pN, with a standard error of 0.7 pN; see Table~\ref{tab:results} for other literature data on EggPC.

Each studied vesicle was permeabilized, which led to partial sugar mixing. If the associated loss of optical contrast was not significant, we used the same vesicle a few more times to perform edge tension measurements on the same vesicle and test for reproducibility. In this case, the pulses were well separated in time, e.g. by an interval of approximately 5 min. The total of 41 experiments were carried out on 16 GUVs from different preparations.


\subsection*{Effect of membrane composition}

Having tested our approach on EggPC with significantly better accuracy as compared to other methods, we proceeded with examining membranes of different compositions. We first tested vesicles made of DOPC because (\textit{i}) this lipid has been used in a few other studies \citep{Gen93,Kar03,Chi06}, which could provide basis for comparison with our results; and (\textit{ii})  DOPC is a pure lipid type, as compared to EggPC which is composed of several lipid species that may influence the measured edge tensions. 

The edge tensions for DOPC membranes obtained from electroporation of black lipid membranes \citep{Gen93}, light-induced poration of GUVs \citep{Kar03} or AFM studies on supported bilayers \citep{Chi06} range from 3.9 to 25 pN, see Table~\ref{tab:results}. The upper value corresponds to DOPC purchased from the same producer as ours (Avanti Polar Lipids); for details on effects of lipids from different producers see \citep{Kar03}. Applied to DOPC GUVs, our method yielded for the edge tension $\gamma$ = 27.7$\pm$2.5\ pN, from a total of 24 experiments. This value is comparable to previous results.

After characterizing the edge tension of pure DOPC membranes, we proceeded with examining differences resulting from the presence of small fractions of guest molecules in the membrane. We considered the effect of cholesterol because it is ubiquitous in eukaryotic membranes. The inverted-cone shape of this molecule should prevent it from locating at the rim of the pore. Thus, the presence of cholesterol would require more energy to rearrange the lipids along the pore walls and the edge tension is expected to increase \citep{Kar03}. Indeed, for membranes made of DOPC:cholesterol in 5:1 molar ratio, we found a mean edge tension value of 36.4 pN with a standard error of 1.9 pN from a total of 14 experiments, confirming the expectations and a trend reported previously \citep{Kar03}.

Similarly to cholesterol, we probed the effect of DOPE, which to our knowledge has not been explored so far. Like cholesterol, this lipid also has an inverted-cone shape and is thus expected to lead to an increase in the edge tension \citep{Kar03}. However, membranes with the same molar fraction of the guest molecules, i.e., DOPC:DOPE with 5:1 molar ratio, yielded quite an unexpected result. We found a significant decrease of the  edge tension, as compared to pure DOPC, with a value of 15.6$\pm$ 1.3 pN as estimated from 31 experiments.

All our results are summarized in Table~\ref{tab:results} including edge tensions obtained by other groups.

\subsection*{Electroporation thresholds \corr{and stability of} membranes with various compositions}

While EggPC, DOPC, and DOPC:DOPE GUVs exposed to electric pulses of a few tens of kV/m porated and resealed very quickly, such field amplitudes completely destabilized DOPC:cholesterol vesicles. The GUVs burst and disintegrated in a fashion reminiscent of that observed with charged membranes \citep{Ris09}. An image sequence of such an event is given in Fig.~\ref{fig:burst}. To avoid vesicle destruction, for the edge tension measurements we were applying pulses with lower amplitudes, around half the strength we used for the other compositions. With these reduced amplitudes, DOPC:cholesterol vesicles porated and resealed in the usual manner. Note that such weak pulses did not porate the cholesterol-free vesicles. This indicates that cholesterol lowers the stability of DOPC membranes when exposed to electric pulses.

As a characteristic of the membrane stability in electric fields one can consider the critical transmembrane potential $\Delta \psi _c$, at which poration occurs. Following \cite{Neu89}, at the moment of maximally expanded pore, we define the critical transmembrane potential as $\Delta \psi _c=(3/2)R_{in} E\cos(\theta _p)$, where $R_{in}$ denotes the initial vesicle radius before poration, $E$ is the field magnitude and $\theta _p$ is the inclination angle defining the location of the pore edge with respect to the vesicle center; \corr{see Supporting Material}. This expression can be presented as $\Delta \psi _c=(3/2)E\sqrt{R_{in}^{2}-r_m^{2}}$, where $r_m$ is the maximal pore size. Thus, measuring $r_m$ after applying a pulse with a field strength $E$ provides us with an estimate for the critical transmembrane poration $\Delta \psi _c$. The results for the four explored mixtures are included in Table~\ref{tab:results}. While the cholesterol-free membranes porate at similar values of $\Delta \psi _c$ around 0.9 V, the addition of 17 mol $\%$ cholesterol to DOPC bilayers (DOPC:cholesterol 5:1) decreases the critical transmembrane potential to around 0.7 V, i.e., destabilizes these membranes when exposed to an electric pulse. 

As mentioned in section Electromediated pore formation, the critical transmembrane potential is related to the so-called lysis tension, $\sigma_{lys}$. \corr{The latter can be deduced, see Table~\ref{tab:results}, provided knowledge about the membrane thickness is available; see Supporting Material for details. From the decrease in the critical transmembrane potential we find that cholesterol lowers the lysis tension of DOPC membranes.}





\section*{Discussion}

In this section, we will compare previously reported results for the edge tension obtained from other methods and data measured here. Before that, we emphasize some advantages of our approach. First, it is relatively easy to apply in a moderately equipped lab. The experiments can be performed using a simple home-made chamber with two electrodes and a microscope. Even though we used a fast digital camera for the acquisition of the images, we did not exploit the full potential of this equipment. Indeed, conventional cameras with acquisition speed of around 1000 fps are now cheap and widely available on the market. Such acquisition speed is more than sufficient for collecting reasonable amount of data in the time range of interest between 5 and 100 ms; see Fig.~\ref{fig:figure2}. Second, the total measuring time is less than 5 min. This allows us to perform a significant number of measurements and achieve good data statistics. Note that studies reporting measurements on giant vesicles in the literature very often contain not more than 5 experiments (for example, the method reported in \citep{Har79} was applied to 3 vesicles only). Most often, the number of measurements is not even mentioned. Achieving good data statistics is important particularly when multicomponent membranes are examined. For example, in vesicles prepared from DOPC, sphingomyelin and cholesterol, Veatch and Keller \citep{Veatch03} have estimated a variation of ± 2 mol $\%$ cholesterol between individual vesicles prepared by the electroformation method. Inspection of fluorescent images of DOPC, dioleoylphosphatidylglycerol and cholesterol, which have undergone phase separation, has suggested that this spread may be even larger for certain compositions \citep{Vequi10}. For this reason, when dealing with multicomponent vesicles one should examine large populations of liposomes. The growing research activity on more complex membrane compositions accentuates the need of developing methods which can quickly and easilly be applied on many vesicles. Our method fulfills this requirement. Third, compared to other methods, our approach is of high precision and allows measuring the edge tension with a relatively small error. Finally, in contrast to the procedures in \citep{Kar03,Pue03}, our experimental approach does not require the presence of (\textit{i}) a very viscous solvent (typically 66 vol $\%$ glycerol) to slow down the pore closure dynamics and (\textit{ii}) fluorescent dye to visualize the pore. As discussed above, both of these compounds may influence the edge tension value.

We measured the edge tension of membranes with different compositions, and found it to be 27.7 and 14.2 pN for pure DOPC and EggPC bilayers, respectively. Considering the multicomponent character of EggPC, this result is understandable: small fractions of edge active lipid molecules in EggPC can stabilize the pore and lower the edge energy as compared to that of pure DOPC bilayers. Our literature survey shows a significant scatter of the data for EggPC. Values of $\gamma$ were reported to range from 8.6 to 42 pN; see Table~\ref{tab:results}. Presumably, EggPC purchased from different producers can vary in terms of composition. Impurities in synthetic lipids like DOPC can also lead to significant variation as demonstrated in \citep{Kar03}. Indeed, our results for DOPC are closer to the upper value reported in \citep{Kar03} and to the result measured in \citep{Gen93}, which is to be expected having in mind that our lipid was purchased from the same producer; see Table~\ref{tab:results}. This finding contradicts results obtained from a method based on puncturing supported lipid bilayers by means of an AFM cantilever tip \citep{Loi02}. This latter method seems to give systematically lower values for the edge tension; see Table~\ref{tab:results} for the data on DOPC in \citep{Chi06} and palmitoyloleoylphosphatidylcholine (POPC) in \citep{Gar07}. We can only speculate that the presence of the substrate supporting the bilayer might influence the pore stability and thus the edge tension. Presumably, this method is still applicable when examining relative changes in the edge tension resulting from the presence of edge active molecules \citep{Gar07} or in lipid mixtures \citep{Chi06}.

The inclusion of cholesterol in DOPC membranes (5:1 mol:mol or $\sim$ 17 mol \%) led to two interesting observations: membrane destabilization when exposed to electric pulses and increase in the edge tension. We first discuss the former. 
 
Electric pulses of 5 ms duration inducing transmembrane voltages of around 1.25 V lead to irreversible collapse of the cholesterol-containing liposomes; see Fig.~\ref{fig:burst}. The reversible poration of the liposomes required the application of pulses with about half the field strength, i.e. transmembrane voltages of approximately 0.7 V. Such pulses did not appear to cause poration on other membrane types, but were able to successfully porate the DOPC:cholesterol GUVs. As shown in Table~\ref{tab:results}, the critical transmembrane potential needed to rupture the DOPC membrane is lowered by the presence of cholesterol. \corr{This observation is interesting and unexpected having in mind the somewhat opposite effect of cholesterol observed on stearoyloleoylphosphatidylcholine (SOPC) giant vesicles \citep{Nee89,Zhe93}. In SOPC membranes, cholesterol was found to increase the critical poration threshold $\Delta \psi _c$, see Table~\ref{tab:results}. On the contrary, dipalmitoylphosphatidylcholine (DPPC) vesicles doped with 12 and 20 mol$\%$ of cholesterol appear to porate at transmembrane potentials lower than that of pure DPPC membranes \citep{Raf99}. Finally, for EggPC planar membranes \citep{Gen93} and vesicles \citep{Raf99} no effect of cholesterol on the critical permeabilization potential was observed. In summary, cholesterol, which alters the hydrophobic core of the bilayer, affects the membrane stability in a different fashion depending on the molecular architecture of the lipid building the membrane.}

Recent studies suggest that the effect of cholesterol on another material property of membranes, namely the bending stiffness, is also lipid specific and depends on the degree of unsaturation of the acyl chains \citep{PanPRL08,PanBJ08,Gra09}. Earlier conventional beliefs were that by ordering the acyl chains in fluid membranes, cholesterol inherently leads to an increase in their bending rigidity. This concept was supported by observations on lipids such as SOPC \citep{Evans90,Song93}, dimyristoylphosphatidylcholine (DMPC) \citep{Duwe90,Meleard97} and POPC \citep{Henriksen06}. However, as demonstrated recently \citep{PanPRL08,Gra09}, the bending rigidity of DOPC-cholesterol mixtures does not show any significant correlation with the cholesterol content, while in sphingomyelin-cholesterol membranes it even decreased \citep{Gra09}. \corr{The origin of the cholesterol-induced lowering of the critical poration potential in DOPC membranes could be sought in its effect on decreasing the membrane conductivity justified in detail in \citep{Raf99}. This behavior may be related to lipid ordering. Our results suggest that the influence of cholesterol on the mechanical properties of lipid membranes depends on the specific lipid architecture and is far from being fully elucidated.} Further studies of the influence of cholesterol on the membrane properties as a function of lipid type and cholesterol content have still to be performed. Here, we will only emphasize that the critical transmembrane potential of such membranes is a physical parameter very sensitive to the membrane composition and can be employed to characterize various types of membranes and the effect of cholesterol as shown by our data.


While DOPC membranes appear to be destabilized by the addition of cholesterol, as illustrated by a decrease in the critical transmembrane potential \corr{and the lysis tension}, the edge tension is observed to increase from 27.7 pN for pure DOPC to 36.4 pN for vesicles made of DOPC:cholesterol 5:1. This is consistent with the trend reported in \citep{Kar03} where the dependence of the edge tension $\gamma$ as a function of the cholesterol molar fraction $x$ was empirically extracted to be $\gamma =\gamma _0+26.7x$, where $\gamma _0$ is the edge tension of the cholesterol-free bilayer. This expression predicts slightly lower edge tension both for the DOPC membranes measured here and for SOPC vesicles studied in \citep{Zhe93}; see Table~\ref{tab:results} for the exact values. The effect of cholesterol increasing the edge tension is expected, considering the inverted-cone shape of the molecule which is supposed to penalize lipid rearrangement at the pore edge \citep{Kar03}.

To summarize, pore stability in DOPC membranes is less favored in the presence of cholesterol. The larger value of $\gamma$ makes pores close faster. On the other hand, cholesterol reduces the lysis tension by inducing structural changes in the bilayer. This is corroborated by our observation that weaker pulses, which do not have any significant effect on cholesterol-free vesicles were sufficient to porate the DOPC:cholesterol liposomes.

We also studied the edge tension effect of another lipid molecule, DOPE, with the same fraction of approximately 17 mol \% in DOPC GUVs. Compared to pure DOPC membranes, we found a significant decrease in the edge tension, from 27.7 to 15.6 pN upon the addition of DOPE, reminding of effects induced by surfactants \citep{Kar03,Pue03}; see the value of $\gamma$ for DOPC mixtures with the detergent Tween 20 in Table~\ref{tab:results}. Interestingly, similar values as those for $\gamma$ of DOPC:DOPE 5:1 were obtained for EggPE and PE extracts from Escherichia coli \citep{Che85,Mel90}. Within the framework of the membrane elasticity theory and considering the inverted-cone shape of the DOPE molecule, one would expect an increase in the edge tension upon DOPE addition, analogously to the effect of cholesterol \citep{Kar03}. Our results contradict this view. \corr{Presumably, the molecular architecture of PE-lipids leading to the tendency to form inverted hexagonal phase, which facilitates fusion and vesicle leakage, see e.g. \cite{Ellens86}, is also responsible for stabilizing pores and lowering the edge tension in porated membranes as demonstrated here. A plausible explanation for this behavior is also provided by the propensity of PE to form interlipid hydrogen bonds \citep{Lewis93,Pink98}. This behavior is not observed for pure PC bilayers where the edge tension is higher. Our observations suggest that simplistic pictures based on the inverted-cone shape of DOPE are not realistic for the interpretation of pore behavior and edge tensions. Inter-PE hydrogen bonding in the pore region can effectively stabilize pores. It would be interesting to test this hypothesis with simulation studies.}


\section*{Conclusion}
We have successfully developed and applied a new method for edge tension measurements on GUVs. This method is based on the theoretical model introduced in \citep{Bro00} and relies on the study of pore closure dynamics. Combining membrane electroporation, fast phase contrast imaging and digital image analysis, we have been able to monitor the rate of pore closure, and deduce the edge tension of membranes with different compositions. This method is fast and quite easy to apply, and we believe it could be of use to researchers interested in the mechanical properties of biological membranes. We confirmed trends reporting the tendency of cholesterol to increase the edge tension of lipid membranes. However, cholesterol appears to have an unconventional and lipid-specific effect on the membrane stability in electric fields as reflected by the values measured for the critical transmembrane potential and lysis tension. Surprisingly, we found that the addition of DOPE in DOPC vesicles leads to a decrease in the edge tension, which is not, even qualitatively, predicted by the current understanding in the field. We hope our result will provoke some theoretical work in this direction. 


\section*{Supporting material}
Text and one figure are available at http://www.biophysj.org

\section*{Acknowledgements}
TP's stay in Germany was supported by a grant for international mobility of PhD students from the Universit\'e de Toulouse. We thank J.~Teissi\'e for his assistance in purchasing the $\beta$tech pulse generator, and M.-P.~Rols and D.~S.~Dean for critical reading of the manuscript. We acknowledge financial support from the Association Fran\c{c}aise contre les
Myopathies.


\bibliography{biblio}

\clearpage
\section*{Table and figures}

\clearpage

\clearpage

\clearpage
\begin{sidewaystable}[htbp]
\tiny
\begin{tabular}{cccccc}
\textbf{Membrane composition} & \textbf{Edge tension,} & \textbf{Method} & \textbf{Poration threshold,} & \textbf{Lysis tension} &  \textbf{Reference} \\
 &  \textbf{$\gamma$, in pN} &  & \textbf{$\Delta \psi _c$, in mV} & \textbf{$\sigma_{lys}$, in mN/m} & \\
\hline
\\
EggPC & 14.2 $\pm$ 0.7 [41]& Pore closure dynamics and electroporation& 920 $\pm$ 20 & 3.79 & This work \\
EggPC & 20 & Observation of open cylindrical vesicles & & & \citep{Har79} \\
EggPC & 8.6 $\pm$  0.4 & Electroporation of black lipid membranes & & & \citep{Che85} \\
EggPC & 42 & Observation of the disk-vesicle transition & & & \citep{Fro86} \\
EggPC & 8.6 $\pm$ 0.4 & Electroporation of black lipid membranes & & & \citep{Mel90} \\
EggPC & 21 & Electroporation of black lipid membranes & & & \citep{Gen93} \\
\\
DOPC & 27.7 $\pm$ 2.5 [24]& Pore closure dynamics and electroporation & 950 $\pm$ 30 & 4.08 & This work \\
DOPC & 25 & Electroporation of black lipid membranes & & & \citep{Gen93} \\
DOPC from Avanti & 20.7 $\pm$ 3.5 & Pore closure dynamics and light-induced poration & &  & \citep{Kar03} \\
DOPC from Sigma & 6.9 $\pm$ 0.42 & Pore closure dynamics and light-induced poration &  & & \citep{Kar03} \\
DOPC & 13--18 & Pore closure dynamics and light-induced poration & &  & \citep{Pue03} \\
DOPC & 3.9 $\pm$ 0.3 & Atomic force microscopy & & & \citep{Chi06} \\
\\
DOPC:cholesterol 5:1 ($mol$) & 36.4 $\pm$ 1.9 [14]& Pore closure dynamics and electroporation & 680 $\pm$ 20 & 1.89 & This work \\
DOPC:cholesterol mixtures & 9--22 & Pore closure dynamics and light-induced poration & & &  \citep{Kar03} \\
SOPC & 9.2 $\pm$ 0.7 & Micropipette aspiration and electroporation & 1100 & 4.91 &\citep{Zhe93,Nee89} \\
SOPC/cholesterol 1:1 ($mol$) & 30.5 $\pm$ 1.2 & Micropipette aspiration and electroporation & 1800 & & \citep{Zhe93,Nee89} \\
SOPC/cholesterol 1:1 ($mol$) & 26 & Reinterpretation of results from \citep{Zhe93} &  & & \citep{Mor97} \\
\\
DOPC:DOPE 5:1 ($mol$) & 15.6 $\pm$ 1.3 [31]& Pore closure dynamics and electroporation & 880 $\pm$ 40 &  &  This work \\
EggPE & 15 $\pm$ 1 & Electroporation of black lipid membranes & &  &  \citep{Mel90} \\
Escherichia coli PE & 16 $\pm$  0.6 & Electroporation of black lipid membranes & &  &  \citep{Che85} \\
DOPC:Tween 20 mixtures & 0.2--12 & Pore closure dynamics and light-induced poration & &  &  \citep{Kar03,Pue03} \\
\end{tabular}
\caption{Edge tensions for different membrane compositions as measured here and by other groups. The standard errors are given, except if not indicated in the respective literature reference. The numbers in the square brackets indicate the measurements performed for each composition. Data on the critical transmembrane potential \corr{and estimates for the lysis tension, see Supporting Material,} are also given.}
\label{tab:results}
\end{sidewaystable}

\end{document}